\documentstyle[11pt]{article}
\begin{document}
\def\beq{\begin{equation}}
\def\eeq{\end{equation}}
\def\beqa{\begin{eqnarray}}
\def\eeqa{\end{eqnarray}}
\input b0.tex
\input b1.tex
\input b2.tex
\input b3.tex
\input b4.tex
\input bref.tex
\end{document}
\begin{titlepage}
\vspace*{-1cm}
\noindent
\phantom{bla}
\hfill{${UMHEP-411 \atop UH-511-800-94}$}\\
\vskip 2.5cm
\begin{center}
{\Large\bf Uncertainties from Long Range Effects \\
in $B \to K^* \gamma$} \\
\end{center}
\vskip 1.6cm
\begin{center}
{\large Eugene Golowich$^{(a)}$ and Sandip Pakvasa$^{(b)}$} \\
\vskip 0.3cm
$^{(a)}$Department of Physics and Astronomy \\
University of Massachusetts, Amherst MA 01003, USA\\
$^{(b)}$Department of Physics and Astronomy \\
University of Hawaii, Honolulu HI 96822, USA \\
\vskip 0.3cm
\end{center}
\vskip 2cm
\begin{abstract}
\noindent
We reconsider the `long-range' component of the radiative
transition $B \to K^* \gamma$.  A careful analysis of the
vector dominance amplitude $B \to V_1 V_2 \to V_1 \gamma$ is
carried out, with emphasis on the role of gauge invariance.
The procedure for incorporating phenomenological $B\to V_1 V_2$ data
is identified and polarization data, only recently available,
is employed to estimate the magnitude of the vector dominance
effect.  We summarize uncertainties in the $B\to K^*\gamma$ radiative
transition produced by long-range effects and provide sugggestions
for further experimental work.
\end{abstract}
\vfill
\end{titlepage}
\vskip2truecm
\section{\bf Introduction}

Some time ago, we considered the possibility that the
flavor-changing radiative transition $B \to K^* \gamma$ might
experience a contribution from a so-called `long-range'
component.$^{\cite{gp}}$  Among the possible contributions studied were
those in Fig.~1.  We concluded that the vector-meson-dominance (VMD)
diagram of Fig.~1(a) was most likely the largest such contributor,
with the dominant process being $B \to K^* \Psi \to K^* \gamma$.
We then estimated the relative magnitude of the
short-distance and VMD amplitudes.  Since data on
exclusive hadronic $B$ decays was practically nonexistent
at that time, we used theoretical estimates for the VMD amplitude.
Part of the motivation for this paper is to update our original
analysis in light of today's improved database.  In addition,
we wish to present a careful justification for using the VMD concept
in flavor-changing radiative decays and also to explicitly show how
phenomenology of the decay $B \to V_1 V_2$ can be adopted via the
VMD process to the radiative transition $B \to V \gamma$.  Hopefully,
this will clarify some confusion on this subject.
\phantom{xxxx}\vspace{0.1in}
\begin{center}
\begin{tabular}{c}\phantom{xxxxxxxxxxxxxxxxxxxxxx} \\
\phantom{xxxxxxxxxxxxxxxxxxxxxx} \\ \hline
\phantom{xxxxxxxxxxxxxxxxxxxxxx} \\
\phantom{xxxxxxxxxxxxxxxxxxxxxx} \\
\phantom{xxxxxxxxxxxxxxxxxxxxxx} \\
\phantom{xxxxxxxxxxxxxxxxxxxxxx} \\ \hline
\phantom{xxxxxxxxxxxxxxxxxxxxxx} \\
\phantom{xxxxxxxxxxxxxxxxxxxxxx} \\
{Figure 1. Long-range effects} \\
\end{tabular}
\end{center}
\vspace{0.1in}

It is well-chronicled in the literature just how active the study of
the $B \to K^* \gamma$ mode has become, particularly with
the recent experimental detection of this mode,$^{\cite{cleo1}}$
\beq
{\rm Br}_{B \to K^* \gamma} = (4.5 \pm 1.5 \pm 0.9)\times 10^{-5} \ \ .
\label{intro1}
\eeq
As is well known, this determination is in accord
(within errors) with expectations of the Standard Model prediction
based upon the `short-distance' electromagnetic penguin transition.
Of course, marked progress in lowering the present
$39\%$ uncertainty in the observed signal is anticipated.
As this happens, an increased burden will fall upon theorists to
properly interpret the experimental finding.

Although there is qualitative agreement regarding the importance
of the $EM$-penguin effect, the current theoretical situation
is far from resolved in at least two respects.  As pointed out
by Buras and co-workers$^{\cite{buras}}$, scale-dependence
occurring at leading-order ($LO$) in $QCD$ radiative
corrections produces an uncertainty at the $25\%$ level in the
inclusive branching ratio ${\rm Br}_{b \to s \gamma}$.  The
dependence on scale would be reduced in a complete
next-to-leading order ($NLO$) calculation, but analysis at this
level has yet to be completed and formidable calculational
complexities lie ahead.

Moreover, precise determination of the ratio $\Gamma_{B\to K^* \gamma} /
\Gamma_{b \to s \gamma}$ of exclusive to inclusive decay
rates is still a somewhat controversial subject.  Presumably the very
recent CLEO result,$^{\cite{patt}}$
\beq
{\rm Br}_{b \to s \gamma} = (2.32 \pm 0.51 \pm
0.29 \pm 0.32)\times 10^{-4} \ \ ,
\label{intro2}
\eeq
which combined with Eq.~(\ref{intro1}) implies
\beq
{{\rm Br}_{B \to K^* \gamma} \over {\rm Br}_{b \to s \gamma}} =
0.194 \pm 0.094 \ \ ,
\label{intro3}
\eeq
will begin the process of finally resolving this issue.  The range of
theoretical predictions, spanning almost two orders-of-magnitude, which
appear in the literature is distressingly large.  The situation is
perhaps not surprising in view of the array of methods employed, from
potential models to lattice-theoretic simulations.  Although it is
encouraging that the spread in lattice-based estimates is not as
large, recent results ranging from $6\%$ to $23\%$ indicate that more
work is needed.$^{\cite{soni},\cite{bowler},\cite{martinelli}}$

Despite the above uncertainties, it is clear that studies of the
$B \to K^* \gamma$ decay have attained an impressive level of maturity.
We have every reason to expect that the physics of this reaction
will ultimately be understood.  We feel that part of this understanding
should involve the role of long-range effects.
Let us now summarize the contents to follow.  In Sect.~2, we address
the VMD effect by analyzing the topic of vector-vector final
states in $B$ meson decay ($B\to V_1 V_2$) and
the vector meson - photon ($V\gamma$) conversion process.
In particular, we show how to extract relevant information from
the $B\to V_1 V_2$ amplitude and we also review the current status of
the database.  Then in Sect.~3, we consider other possible
long-range contributions such as pole diagrams, which are induced
by the weak mixing of pseudoscalar and/or vector $B$ mesons
with non-$b$-flavored states.  Our conclusions and recommendations for
future study are given in Sect.~4.

\section{\bf Vector Dominance Amplitude}

We wish to consider long range contributions to the transition
\beq
B({\rm p}) \to K^* ({\bf k}, \lambda ) + \gamma ({\bf q}, \sigma ) \ \ .
\label{a1}
\eeq
The transition amplitude can be written in gauge invariant form as
\beqa
\lefteqn{{\cal A}_{B\to K^* \gamma}  = \epsilon_\mu^\dagger
(k ,\lambda ) \epsilon_\nu^{\dagger} (q, \sigma) } \nonumber \\
& & \times \left[ {\overline B} \left( p^\mu p^\nu - g^{\mu\nu} q
\cdot p \right) + i{\overline C} \epsilon^{\mu\nu\alpha\beta} k_{\alpha}
p_\beta \right] \ ,
\label{a2}
\eeqa
where the superbars denote working in the $B$ rest frame for the
$B \to K^* \gamma$ process.  Note the presence of the two independent
amplitudes ${\overline B}$ and ${\overline C}$, which carry the
dimension of inverse energy and are respectively parity-violating and
parity-conserving.  In general, both amplitudes are required because
the weak interaction does not respect parity invariance.$^{\cite{comm1}}$

The $B \to K^* \gamma$ decay rate is given by
\beq
\Gamma_{B \to K^* \gamma} = {|{\bf q}|^3 \over 4\pi} ~\cdot~
\left[ |{\overline B}|^2 + |{\overline C}|^2 \right] \ \ ,
\label{a4}
\eeq
where ${\bf q}$ is the decay momentum in the $B$ rest frame,
\beq
|{\bf q}| = {m_{\rm B}^2 - m_{K^*}^2 \over 2 m_{\rm B}} \ \ .
\label{a5}
\eeq
The branching ratio of Eq.~(\ref{intro1}) together with the average
$B$ lifetime value$^{\cite{patt},\cite{rou}}$
\beq
\tau_B = (1.63 \pm 0.07) \times 10^{-12} {\rm sec.} \ \ ,
\label{a6}
\eeq
implies that the transition amplitude has magnitude
\beqa
\bigg| {\cal A}_{B\to K^* \gamma}^{\rm (expt)} \bigg|  &\equiv&
\sqrt{|{\overline B}|^2 + |{\overline C}|^2} \nonumber \\
&=& \left[ {4 \pi \Gamma_{B \to K^* \gamma} \over |{\bf q}|^3} \right]^{1/2}
\nonumber \\
&=& (3.68 \pm 0.72) \times 10^{-9}~{\rm GeV}^{-1} \ \ .
\label{a6a}
\eeqa
A branching ratio determination alone does not distinguish between the
parity-conserving and parity-violating amplitudes.  Polarization
data is required to disentangle them.

\begin{center}
{\bf The $B \to V_1 V_2$ Transition}
\end{center}

Application of the VMD concept to flavor-changing radiative decays is a
subtle issue. This is partly because VMD is not a basic tenet of the
Standard Model.  That is, it is not explicitly present in the
fundamental description of how quarks couple to gluons or to the
electroweak gauge bosons nor is it associated with the Higgs sector.
Rather it is a product of phenomenology, having been originally motivated
by the similarity between photon-hadron scattering processes and purely
hadronic reactions.  Actually, this original application of the VMD
concept resembles its proposed use here in weak radiative decays since
the photon-hadron scattering and radiative decays both involve physical
external-leg photons.$^{\cite{gsw}}$  At any rate, the task is to formulate
a radiative transition amplitude which respects basic principles and
which utilizes VMD parameters (the $\{ f_{{\rm V}} \}$) which are determined
from $V \to \ell^+ \ell^-$ data.

Although we are primarily concerned with the decay $B \to K^* \gamma$,
much of what follows is true for a more general transition
$B\to M \gamma$, where $M$ is a meson of nonzero spin.
The VMD contribution to such a general flavor-changing radiative decay
is depicted in Fig.~2, where (i) the pseudoscalar meson $B$
decays weakly into meson $M$ and a virtual neutral vector meson $V$,
followed by (ii) the electromagnetic VMD conversion of $V$ into a
photon.  There are two main issues, whether such a VMD amplitude is
`really there' and if so, how to properly use $B\to MV$ data as input.
\phantom{xxxx}\vspace{0.1in}
\begin{center}
\begin{tabular}{c}\phantom{xxxxxxxxxxxxxxxxxxxxxx} \\
\phantom{xxxxxxxxxxxxxxxxxxxxxx} \\ \hline
\phantom{xxxxxxxxxxxxxxxxxxxxxx} \\
\phantom{xxxxxxxxxxxxxxxxxxxxxx} \\
\phantom{xxxxxxxxxxxxxxxxxxxxxx} \\
\phantom{xxxxxxxxxxxxxxxxxxxxxx} \\ \hline
\phantom{xxxxxxxxxxxxxxxxxxxxxx} \\
\phantom{xxxxxxxxxxxxxxxxxxxxxx} \\
{Figure 2. VMD Amplitude for $B \to M + V$} \\
\end{tabular}
\end{center}
\vspace{0.1in}

Consider the decay $B\to V_1 V_2$ of the $B$ into two
vector mesons, where $V_2$ is electrically neutral.  The
constraint of angular momentum conservation states
\beq
{\bf J}_B = {\bf J}_{V_1 V_2} \qquad {\rm with} \qquad
{\bf J}_{V_1 V_2} = {\bf L} + {\bf S} \ \ ,
\label{ang}
\eeq
where ${\bf S}$ and ${\bf L}$ are the $V_1 V_2$ total spin
and orbital angular momentum.
Since meson $B$ is spinless, we have ${\bf J}_B = 0$ and so
the three possible total spins $S =0,1,2$ of the
$V_1 V_2$ state must be accompanied by the three
orbital angular momenta $L = 0,1,2$.  Thus there are three
independent amplitudes ${\cal M}^{(\ell )}\ (\ell = 0,1,2)$.
One could instead use helicity amplitudes
${\cal M}_{\lambda_1 \lambda_2}$. As indicated in Figure~3, these
correspond to the three independent choices $\lambda_1 \lambda_2 =
++, --, 00$.  Throughout this paper, we shall classify these three
configurations either as `transverse' $T$ (for $++, --$) or
`longitudinal' $L$ (for $00$).
\phantom{xxxx}\vspace{0.1in}
\begin{center}
\begin{tabular}{c}\phantom{xxxxxxxxxxxxxxxxxxxxxx} \\
\phantom{xxxxxxxxxxxxxxxxxxxxxx} \\ \hline
\phantom{xxxxxxxxxxxxxxxxxxxxxx} \\
\phantom{xxxxxxxxxxxxxxxxxxxxxx} \\
\phantom{xxxxxxxxxxxxxxxxxxxxxx} \\
\phantom{xxxxxxxxxxxxxxxxxxxxxx} \\ \hline
\phantom{xxxxxxxxxxxxxxxxxxxxxx} \\
\phantom{xxxxxxxxxxxxxxxxxxxxxx} \\
{Figure 3. Helicity Configurations in $B \to V_1 V_2$} \\
\end{tabular}
\end{center}
\vspace{0.1in}

Calculations are most easily performed in terms of
Lorentz-covariant kinematic variables and their corresponding
invariant amplitudes.  Following Valencia,$^{\cite{val}}$
one can denote the three invariant amplitudes as $a,b,c$ and write
\beqa
\lefteqn{{\cal M}_{\lambda_1 \lambda_2} = \
_{\rm out}\langle V_1 (k_1,\lambda_1) V_2 (k_2,\lambda_2)|
B(p)\rangle_{\rm in}} \nonumber \\
& & = \epsilon_\mu^\dagger (k_1 ,\lambda_1)
\epsilon_\nu^{\dagger} (k_2 ,\lambda_2)
 \left[ a g^{\mu\nu} + {b \over m_1 m_2} p^\mu p^\nu +
i{c\over m_1 m_2} \epsilon^{\mu\nu\alpha\beta} k_{1\alpha} p_\beta
\right] .
\label{v1}
\eeqa
Note that the $a$, $b$, $c$ amplitudes each carries the dimension of
energy.  The set of helicity amplitudes is constructed by writing
explicit representations of the $V_1 , V_2$ polarization
vectors.  We summarize the results here for convenience,
\beqa
{\cal M}_{++} &=&  a - \sqrt{x^2 - 1}~c \nonumber \\
{\cal M}_{--} &=&  a + \sqrt{x^2 - 1}~c  \label{v1a} \\
{\cal M}_{00} &=& - x~a + (x^2 - 1)~b  \ \ , \nonumber
\eeqa
where $x$ is defined by
\beq
x \equiv {k_1 \cdot k_2 \over m_1 m_2} =
{ m_{\rm B}^2 - m_1^2 - m_2^2 \over 2m_1 m_2}
\label{v9}
\eeq
and obeys
\beq
x^2 = 1 + { m_{\rm B}^2 |{\bf k}|^2 \over m_1^2 m_2^2} \ \ .
\label{v10}
\eeq
In Ref.~\cite{val}, amplitude `$c$' is called the P-wave
amplitude, while `$a$' and `$b$' are the S-wave and D-wave
amplitudes.  This nomenclature becomes apparent from
the decay rate for $B \to V_1 V_2$,
\beq
\Gamma_{B \to V_1 V_2} = {|{\bf k}| \over 8\pi m_{\rm B}^2 } \left[
2 |a|^2 + |xa + (x^2 - 1)b|^2 + 2(x^2 - 1)|c|^2  \right] \ \ ,
\label{v8}
\eeq
upon using Eq.~(\ref{v10}).
The $|{\bf k}|$, $|{\bf k}|^3$ and $|{\bf k}|^5$ dependences of
the $a$, $c$ and $b$ amplitudes mirror the threshold behaviors expected
of S-waves, P-waves and D-waves respectively.  Decay rates corresponding
to helicity configurations are obtained
from the helicity amplitudes of Eq.~(\ref{v1a}),
\beq
\Gamma_{B \to V_1 V_2} = {|{\bf k}| \over 8\pi m_{\rm B}^2} \left[
|{\cal M}_{++}|^2 + |{\cal M}_{--}|^2 + |{\cal M}_{00}|^2
\right] \ \ .
\label{v8a}
\eeq
As expected, the various helicity contributions are decoupled since
they are physically distinct.

\begin{center}
{\bf The VMD Amplitude}
\end{center}

Let us consider the construction of a VMD amplitude using
$B\to V_1 V_2$ as input.  There are problems with taking a
theoretical model for $B\to V_1 V_2$ since even if the model is
arranged to fit the $B \to V_1 V_2$ transition rate, the vector meson
polarizations may well not agree with experiment.$^{\cite{dtp}}$
Therefore, we adopt a phenomenological approach.
Suppose all the invariant amplitudes $a,b,c$ have been determined in
terms of experimental data from $B\to V_1 V_2$ measurements.
The next step is then to continue the $B\to V_1 V_2$ decay amplitude from
$k_2^2 = m_2^2$ to $k_2^2 = 0$ such that the meson $V_2$ propagates as a
massless virtual particle before converting to a photon.  Throughout,
however, the mass parameter $m_2$ in Eq.~(\ref{v1}) will remain fixed at its
physical value since it is present in the definition of amplitudes $b$, $c$
simply for dimensional reasons.
Using the {\it full} $B\to V_1 V_2$ amplitude in the VMD calculation
results in the following $B \to V_1 \gamma$ amplitude,
\beqa
\lefteqn{{\cal A}_{\rm VMD} = {e \over {\overline f}_{\rm V}}
\epsilon_\mu^\dagger (k ,\lambda) \epsilon_\nu^{\dagger} (q ,\sigma)}
 \nonumber \\
& & \times \left[ {\overline a} g^{\mu\nu} + {{\overline b}
\over m_1 m_2} p^\mu p^\nu + i{{\overline c}\over m_1 m_2}
\epsilon^{\mu\nu\alpha\beta} k_{\alpha} p_\beta \right] \ .
\label{v2}
\eeqa
where we denote the kinematics of the radiative process with
superbars and take $k_1 \to k$, $k_2 \to q$ as well.  The
quantity ${\overline f}_{\rm V}$, whose determination we shall discuss
shortly, is the continuation to $k_2^2 = 0$ of a quantity occurring
in the $V_2 \gamma$ conversion vertex.

Then, under a gauge transformation as implemented by $\epsilon^{\nu}
\to q^\nu$, the amplitude of Eq.~(\ref{v2}) responds as
\beq
{\cal A}_{\rm VMD} \to {e \over {\overline f}_{\rm V}} \left[
{\overline a}~ \epsilon_1{\dagger} \cdot q
+ {{\overline b} \over m_1 m_2} q \cdot p~\epsilon_1{\dagger} \cdot p
\right] \ .
\label{v3}
\eeq
The new term must vanish if gauge invariance is to be
maintained.  Upon noting
\beq
p = k + q \quad \Longrightarrow \quad \epsilon_1{\dagger} \cdot p =
\epsilon_1{\dagger} \cdot q\ \ ,
\label{v4}
\eeq
we obtain the following constraint,
\beq
{\overline a} + {{\overline b} \over m_1 m_2} q \cdot p = 0 \ \ .
\label{v5}
\eeq
This condition means that we cannot use the full set of invariant
functions $a,b,c$ in the VMD amplitude for $B \to V_1 \gamma$.
The combination appearing in Eq.~(\ref{v5}) must be avoided.

There is a simple physical interpretation of the above rule.
Consider decay of the pseudoscalar $B$ into
two longitudinally-polarized vector mesons.  Expressed in terms
of the invariant amplitudes from Eq.~(\ref{v1}), the
corresponding amplitude is
\beq
{\cal M}_{00} = -{ 1 \over m_1 m_2} \left[
( E_1 E_2 + {\bf k}^2 )a + {m_{\rm B}^2 {\bf k}^2 \over m_1 m_2} b
\right] \ \ .
\label{v6}
\eeq
In the $k_2^2 \to 0$ limit relevant to the VMD amplitude,
we can re-express the above as
\beq
{\cal M}_{00} = -{ 1 \over m_1 m_2} q \cdot p \left[
{\overline a} + {{\overline b} \over m_1 m_2} q \cdot p \right] \ \ .
\label{v7}
\eeq
Thus, the condition obtained in Eq.~(\ref{v5}) from gauge
invariance is equivalent to demanding that a vanishing contribution
to ${\cal A}_{\rm VMD}$ coming from the $k_2^2 = 0$
off-shell extension of ${\cal M}_{00}$. That is,
if the $B \to V_1 V_2$ amplitudes are to be used as input
to a VMD calculation, then one must not use the `$00$' helicity
amplitude --- it must be discarded.  This result is entirely natural
when viewed in terms of vector-meson photon mixing.  The helicity
of a physical photon must have unit magnitude, so conversion from
a vector-meson with helicity zero is forbidden.  As a corollary, it
follows that if the physical $B \to V_1 V_2$ decay consists entirely of the
$00$ helicity mode, then the corresponding VMD amplitude for radiative
decay will vanish.

Next we turn to consideration of the VMD conversion vertex $V_2
\to \gamma$.  It too must be constructed in a manner consistent
with electromagnetic gauge invariance.  This amounts to demanding that
any mixing experienced by a photon propagating with squared-momentum
$q^2$ should not lead to a nonzero mass when the photon is on-shell
($q^2 = 0$).  The simplest possible conversion vertex involves just
the field strength tensors $F_{\mu\nu}$ and $V^{\mu\nu}$ in the
Lorentz invariant combination $F_{\mu\nu} V^{\mu\nu}$ where
\beq
F_{\mu\nu} = \partial_\mu A_\nu - \partial_\nu A_\mu \qquad {\rm and}
\qquad V_{\mu\nu} = \partial_\mu V_\nu - \partial_\nu V_\mu \ .
\label{vmd1a}
\eeq
However, an interaction of this type implies vanishing matrix elements at
$q^2 = 0$ and would rule out application to any process with external
photons.  Among others, it was most notably
Sakurai$^{\cite{sak}}$ who argued that the above relation, although
gauge-invariant, is too restrictive and should be extended to
\beq
{\cal L}_{\rm VMD} = {e\over f_{\rm V}} \left[ {1\over 2}
F_{\mu\nu} V^{\mu\nu} + J_\mu^{{\rm V}}A^\mu \right] \ \ ,
\label{vmd2}
\eeq
where $J_\mu^{{\rm V}}$ is the (conserved) current to which the vector
meson $V$ couples and $f_{\rm V}$ is defined by
\beq
\langle 0 | eJ_{\rm em}^\mu (0) | V (k ,\lambda)\rangle \equiv
{e m_{\rm V}^2 \over f_V} \epsilon^\mu (k,\lambda )\ \ .
\label{vec}
\eeq
The dimensionless quantity $f_V$ is typically
determined in terms of $V\to \ell^+\ell^-$ data,
\beq
\Gamma_{V\to \ell^+\ell^-} = {4\pi\alpha^2 \over 3}
{m_V\over f_V^2} \left[ 1 - {4m_\ell^2 \over m_V^2} \right]^{1/2}
\left[ 1 + {2m_\ell^2 \over m_V^2} \right] \ \ .
\label{vll}
\eeq
Numerical values extracted in this manner are displayed in
Table~1, where the unit of energy is GeV.
\phantom{xxxx}\vspace{0.1in}
\begin{center}
\begin{tabular}{ccccc}
\multicolumn{5}{c}{Table~1. {The Coefficients $f_V$}}\\ \hline\hline
$V$ & $\Gamma_{V\to e^+e^-}$ & $m_V$ & $f_V$ & $e/f_V$ \\ \hline
$\rho^0$ & $6.77\times 10^{-6}$ & $0.768$ & $5.03$ & $0.06$ \\
$\omega^0$ & $6.03\times 10^{-7}$ & $0.782$ & $17.1$ & $0.018$ \\
$\phi^0$ & $1.37\times 10^{-6}$ & $1.019$ & $12.9$ & $0.024$ \\
$\Psi$ & $5.36\times 10^{-6}$ & $3.097$ & $11.3$ & $0.027$ \\
$\Psi'$ & $2.14\times 10^{-6}$ & $3.686$ & $19.6$ & $0.015$ \\
$\Psi''$ & $0.26\times 10^{-6}$ & $3.770$ & $56.9$ & $0.005$ \\ \hline\hline
\end{tabular}
\end{center}
\vspace{0.1in}

It is clear from the above discussion that each  $f_{\rm V}$
is determined at the physical kinematic value of $k^2 =
m^2_{\rm V}$.  In the VMD amplitude, however, the kinematics
is changed to $k^2 = 0$.  In recognition of this, we have denoted
the off-shell extension appearing in the VMD amplitude as
${\overline f}_{\rm V}$.

\begin{center}
{\bf Phenomenology}
\end{center}

To proceed further, we must make a phenomenological determination of the
invariant amplitudes from $B \to V_1 V_2$ data.  Table~2 displays some
relevant branching fractions, taken from a very recent $CLEO$
analysis of $B$ decays.$^{\cite{alam}}$
As Eq.~(\ref{v8}) for the $B \to V_1 V_2$ width reminds us,
measurement of the decay rate yields only partial information.
To extract the amplitudes $a,b,c$
requires, in addition, polarization measurements.
The current status of such data is, to the
best of our knowledge, summarized in Table~3.$^{\cite{alam}}$  Note
that existing data only distinguishes between longitudinal
and transverse polarization in the $V_1 V_2$ final state.
\phantom{xxxx}\vspace{0.1in}
\begin{center}
\begin{tabular}{cc}
\multicolumn{2}{c}{Table~2. {$B\to V_1 V_2$ Branching Fractions}}\\
\hline\hline
Mode & ${\rm Br}_{B\to V_1 V_2}$ \\ \hline
$B^0 \to {\bar K}^{*0}\Psi$ & $(1.69 \pm 0.031 \pm 0.018)\cdot 10^{-3}$ \\
$B^- \to K^{*-}\Psi$ & $(1.78 \pm 0.051 \pm 0.023)\cdot 10^{-3}$ \\
$B^- \to D^{*0}\rho^-$ & $(1.68 \pm 0.21 \pm 0.25 \pm 0.12)\cdot 10^{-2}$ \\
$B^0 \to D^{*+} \rho^-$ & $(0.74 \pm 0.10 \pm 0.14 \pm 0.03)\cdot 10^{-2}$ \\
\hline\hline
\end{tabular}
\end{center}
\vspace{0.1in}

For the remainder of this section, we shall restrict our discussion
to $B \to K^* \Psi$ decay and take into account
only the $\Psi$ intermediate state.  The vector meson $\Psi$ is
the only one for which relevant data is available and at the same time,
gives rise to the largest VMD amplitude.  From Table~2 and
the lifetime value in Eq.~(\ref{a6}), we find
\beqa
\Gamma^{({\rm L})}_{B \to K^* \Psi} &=& (5.60 \pm 1.55) \times
10^{-13}~{\rm MeV} \ \ , \nonumber \\
\Gamma^{({\rm T})}_{B \to K^* \Psi} &=& (1.40 \pm 0.39) \times
10^{-13}~{\rm MeV} \ \ , \label{p4a}
\eeqa
where neutral and charged modes are averaged over.

On the other hand, the theoretical decay rates of transversely (summed
over the $++$ and $--$ helicity configurations) and longitudinally
polarized particles are
\beqa
\Gamma^{({\rm T})}_{B \to K^* \Psi} &=& {|{\bf k}| \over 4\pi m_{\rm B}^2}
{}~\cdot~\left[ |a|^2 + (x^2 - 1)|c|^2 \right] \ \ , \nonumber \\
\Gamma^{({\rm L})}_{B \to K^* \Psi} &=& {|{\bf k}| \over 8\pi m_{\rm B}^2}
{}~\cdot~ | x~a + (x^2 - 1)~b |^2 \ \ .
\label{p4}
\eeqa
Without more detailed polarization data, we cannot determine all
three $a$, $b$, $c$ amplitudes.  Therefore, let us first
assume that the entire decay of transversely polarized particles
comes from the P-wave amplitude $c$.  If so, we have the relation
\beq
{ |c|  \over m_{K^*} m_{\Psi}}
= \left[ { 4 \pi \Gamma^{({\rm T})}_{B \to K^*\Psi} \over
|{\bf k}|^3 }\right]^{1/2} \ \ .
\label{p4b}
\eeq
The other extreme, where transversely polarized particles
arise from the $a$ amplitude, implies
\beq
|a| = m_{\rm B} \left[ { 4 \pi \Gamma^{({\rm T})}_{B \to K^*\Psi}
\over  |{\bf k}| }\right]^{1/2} \ \ .
\label{p5}
\eeq
For the sake of completeness, we note in passing that
the rate for longitudinal polarization then implies a
value for amplitude $b$, up to a phase ambiguity,
\beq
b_\pm = {m^2_{K^*} m^2_{\Psi} \over m_{\rm B}^2 |{\bf k}|^2 } \cdot
\left[ -x~a \pm \sqrt{ {8 \pi
\Gamma^{({\rm L})}_{B \to K^*\Psi} \over |{\bf k}|}}~\right]
\label{p6}
\eeq
\phantom{xxxx}\vspace{0.1in}
\begin{center}
\begin{tabular}{lc}
\multicolumn{2}{c}{Table~3. {Helicity Content of $B\to V_1 V_2$ Transitions}}\\
\hline\hline
Mode & $\Gamma_L /(\Gamma_L + \Gamma_T )$ \\ \hline
$B \to {\bar K}^* \Psi$ & $0.80 \pm 0.08 \pm 0.05$ \\
$B^0 \to D^{*+} \rho^-$ & $0.93 \pm 0.05 \pm 0.05$  \\ \hline\hline
\end{tabular}
\end{center}
\vspace{0.1in}

Finally, we consider the VMD process $B \to K^* \Psi \to
K^* \gamma$.  This involves both implementing the phenomenological
input obtained above and taking account of possible suppression arising
from the off-shell extrapolation procedure.  As regards the former
point, recall that we are constrained by gauge invariance to work only with
the amplitudes for transversely-polarized particles.  Upon comparing
Eq.~(\ref{a2}) with Eq.~(\ref{v2}), we obtain for the scenario of
pure parity-conserving (PC) decay,
\beqa
\bigg| {\cal A}_{\rm VMD}^{\rm (PC)}\bigg|
&\equiv& {\overline C}_{\rm VMD}
= {e \over {\overline f}_{\Psi}} { |{\overline c}|
\over m_{K^*} m_{\Psi}} \nonumber \\
&\simeq& 3.68 \cdot 10^{-10}~{\rm GeV}^{-1}
\ , \label{p1a}
\eeqa
where we have utilized evidence$^{\cite{paul}}$ that the
extrapolation $k_2^2 = m_{\Psi}^2 \to k_2^2 = 0$ introduces
some suppression in the VMD amplitude, {\it i.e.} $f_\Psi
\simeq 1.5{\overline f}_\Psi$.  In magnitude, the ratio of the above
parity-conserving amplitude to that of the empirical amplitude of
Eq.~(\ref{a6a}) is
\beq
\bigg| { {\cal A}_{\rm VMD}^{\rm (PC)} \over
{\cal A}_{B\to K^* \gamma}^{\rm (expt)}} \bigg| \simeq 0.10 \ \ .
\label{p7}
\eeq
Following a similar procedure for the parity-violating (PV)
amplitude yields the result
\beqa
\bigg| {\cal A}_{\rm VMD}^{\rm (PV)}\bigg|
&\equiv& {\overline B}_{\rm VMD}
= {e \over {\overline f}_{\Psi}} { |{\overline a}|
\over m_{\rm B} E_\gamma } \nonumber \\
&\simeq& 2.26 \cdot 10^{-10}~{\rm GeV}^{-1}
\eeqa
which implies
\beq
\bigg| {{\cal A}_{\rm VMD}^{\rm (PV)} \over
{\cal A}_{B\to K^* \gamma}^{\rm (expt)}} \bigg| \simeq 0.06 \ \ .
\label{p9}
\eeq
We shall discuss these findings in Sect.~5.  Next we address the
estimation of other long-range effects in $B\to K^* \Psi$.
\section{\bf Long Distance Pole Contributions}

By `long distance pole contributions', we mean processes of the type
shown in Figs.~1(b)-(c).  Such amplitudes contain three essential
ingredients:
\begin{enumerate}
\item A weak-mixing in which the nonleptonic weak hamiltonian (depicted
by the circled `W') converts the $B$ meson to a non-bottom-flavored
meson,
\item The radiation of a photon, which occurs independently of the
weak mixing,
\item Propagation of an off-shell meson whose flavor depends
on the relative order of the above two items.
\end{enumerate}
In Figs.~1(a)-(b), there are two particles in the final state, a
photon and a meson $M$ which we shall take as $K^*$ or $\rho$.
In principle, however, meson $M$ can have
any spin other than spin zero.  The intermediate state in
Fig.~1(b), denoted as `$P_n$', is some non-bottom-flavored pseudoscalar meson.
The subscript `$n$' indicates that we are to sum over all
pseudoscalar mesons of the appropriate flavor.  The intermediate state in
Fig.~1(c) has an analogous meaning, except now one sums over all
excited mesons $B_n^*$ except for spin zero.

For definiteness, we shall use the effective weak hamiltonian of
Bauer, Stech and Wirbel$^{\cite{bsw}}$.  The transition operator
appropriate for our purposes is
\begin{equation}
{\cal H}^{\rm (eff)}_{\rm w} = -{G_{\rm F} \over \sqrt{2}}
\left[ :a_1 ({\bar u} b')({\bar d}' u ) +
a_2 ({\bar d}' b')({\bar u} u) : \right] \ \ ,
\label{bsw1}
\end{equation}
where the colons denote normal-ordering and
$b' ,  d'$ are the quark-mixed fields
\begin{eqnarray}
b' &=&  b~V_{ub} + s~V_{ub} \ , \nonumber \\
d' &=&  d~V_{ud} + s~V_{us} \ , \\
s' &=&  s~V_{cs} + d~V_{cd} \ \ .  \nonumber
\label{cab}
\end{eqnarray}
In this paper, we take$^{\cite{patt}}$
\beqa
V_{us} &=& 0.22 \nonumber \\
V_{ub} &=& 0.08*V_{cb} = 0.08*0.040 \simeq 0.003  \ \ .
\label{ckm}
\eeqa

The quark fields occur in left-handed combinations, denoted by
\beq
({\bar q}_1 q_2) \equiv {\bar q}_1 \gamma_\mu (1 + \gamma_5) q_2 \ \ ,
\label{left}
\eeq
and $a_1, a_2$ are free parameters determined by fitting to $2$-body
$B$ decays,$^{\cite{nel}}$
\beq
a_1 = 0.98 \pm 0.03 \pm 0.04 \pm 0.09 \ , \qquad
a_2 = 0.25 \pm 0.013 \pm 0.006 \pm 0.02  \ \ .
\label{bsw2}
\eeq
It is important to remember that in the BSW description,
the effective hamiltonian is to be interpreted such as
Fierz reordering is not allowed.  Thus color-mismatched
matrix elements are forbidden.

\begin{center}
{\bf Pole Amplitudes of Type I}
\end{center}
For amplitudes of this type, the weak mixing occurs prior to photon
emission, as in Fig.~1(b).  For $B\to K^* \gamma$, the virtual
particle $P_n$ which propagates will be a kaon or one of its
pseudoscalar recurrences.  Data availability forces us to consider
just the kaon here.  The decay amplitude ${\cal A}_{\rm pole}^{(I)}$
for the transition $B \to \gamma + M$ has the general form
\beq
{\cal A}_{\rm pole}^{(I)} = \sum_n \ g_{M\gamma P_n}\cdot {1\over m_B^2 -
m_{P_n}^2}\cdot  \langle P_n | {\cal H}^{\rm (eff)}_{\rm w} | B \rangle \ \ .
\label{amp1}
\eeq
With Fig.~1(b) as a guide, the notation should be self-evident.\vspace{0.3in}

The calculation of weak-mixing matrix element
$\langle P_n | {\cal H}^{\rm (eff)}_{\rm w} | B \rangle$
of $B$ with the kaon is straightforward in the factorization
approach,
\beq
\langle K | {\cal H}^{\rm (eff)}_{\rm w} | B \rangle \simeq
a_1 V_{ub} V_{us} f_K f_B m_B^2 G_F/\sqrt{2} \ \ .
\label{amp1a}
\eeq
For the decay constant of the kaon, we take
\beq
f_K = 161~{\rm MeV} \ \ .
\label{cf3}
\eeq
The present situation for the decay constant $f_B$
is somewhat problematic in that only theoretical
estimates exist.  These occur in three categories,
lattice theoretic$^{\cite{cb}}$, $QCD$ sum rule$^{\cite{dom}}$ and
quark model.  Estimates fall in the range $104 < f_B ({\rm MeV}) < 229$.
We shall adopt the value
\beq
f_B \simeq f_K \ \ ,
\label{cf5}
\eeq
which is influenced most heavily by the lattice estimates.

The only other ingredient needed is the radiative coupling
constant $g_{K^* K \gamma}$.  This is obtained from data on
the transition $K^*\to K \gamma$,
\beq
\Gamma_{K^*\to K \gamma} = {g_{K^* K \gamma}^2 \over 12\pi}{\bf q}^3 \ \ ,
\label{cf8}
\eeq
where ${\bf q}$ is the decay momentum in the $K^*$ rest frame.  We find
\beq
g_{K^* K \gamma} = 0.318~{\rm GeV}^{-1} \ \ ,
\label{cf9}
\eeq
where we have taken the average of the $K^{*-}$ and ${\bar K}^{*0}$
radiative decays.  We note in passing the rather large difference ($40\%$)
between the charged and neutral $K^*$ modes.  The type-I poles would be
a possible source of isospin splitting in the $B^{-}$ and
${\bar B}^{0}$ radiative decays, were such an effect detected.

Substituting in all the above values, we obtain
\beq
{\cal A}_{\rm pole}^{(I)} \simeq 3.77 \times 10^{-11}~{\rm GeV}^{-1}
\label{cf10}
\eeq
or equivalently
\beq
\bigg| {{\cal A}_{\rm pole}^{(I)} \over
{\cal A}_{\rm expt}}\bigg| \simeq 0.01 \ \ .
\label{cf11}
\eeq

\begin{center}
{\bf Pole Amplitudes of Type II}
\end{center}

As mentioned above, a type II pole amplitude is one in which the
electromagnetic transition occurs before the weak mixing,
{\it cf} Fig.~1(c).  Thus, we write
\beq
{\cal A}_{\rm pole}^{(II)}
= \sum_n \ \langle V | {\cal H}^{\rm (eff)}_{\rm w} | B_n^*
\rangle  \cdot {1\over m_{B_n^*}^2 - m_{V}^2}\cdot  g_{B_n^* B\gamma} \ \ .
\label{amp11}
\eeq
For a phenomenological approach, the type II transitions are
less accessible than those of type I due mainly to a scarcity of data.
However, with some theoretical input, we shall be able to consider the
contribution from the $B^* (5325)$ intermediate state in detail.

Finally, there is the weak mixing between $B^*$ and $K^*$,
\beq
\langle K | {\cal H}^{\rm (eff)}_{\rm w} | B \rangle \simeq
a_1 V_{ub} V_{us} g_{{\rm K}^*} g_{{\rm B}^*} G_F/\sqrt{2} \ \ .
\label{amp11a}
\eeq
The `decay constant' for a vector meson $V^b$ is defined by the
matrix element
\beq
\langle 0 | V_\mu ^a (0) | V^b ({\rm p} , \lambda ) \rangle =
\delta^{ab} g_{V^b} \epsilon_\mu ({\rm p} , \lambda ) \ \ .
\label{amp2}
\eeq
For the $K^*$, we use the $SU(3)$ estimate
\beq
g_{K^*} \simeq g_\rho \simeq \sqrt{2} {m_\rho^2 \over f_\rho} \simeq
0.166~{\rm GeV}^2 \ \ ,
\label{amp3}
\eeq
where $f_\rho$ is given in Table~1.  The $B^*$ decay constant is
estimated from the heavy-quark-symmetry relation,
\beq
g_{B^*} = m_B f_B \simeq 0.845~{\rm GeV}^2 \ \ ,
\label{amp4}
\eeq
where we employ the numerical estimate for $f_B$ given in Eq.~(\ref{cf5}).

Although there is not sufficient experimental data to infer the
radiative coupling constant $g_{B^* B \gamma}$ phenomenologically,
this quantity has been estimated in Ref.~\cite{bgw}.  These authors
correctly identify the $B^* \to B \gamma$ decay as a magnetic dipole
transition and thus express the radiative coupling in
terms of the $B^*$ magnetic moment,
\beq
g_{B^* B\gamma} = e \left( {Q_b \over m_b} + {Q_q \over m_q} \right)
\simeq 0.61~{\rm GeV}^{-1} \ \ .
\label{amp4a}
\eeq

Thus we conclude
\beq
{\cal A}_{\rm pole}^{(II)} \simeq 1.79 \times 10^{-11}~{\rm GeV}^{-1} \ \ ,
\label{amp5}
\eeq
which is roughly half the size of the type-I amplitude.  What about
higher $B^*$ excitations?  A consequence of the BSW approach is that
only states with $J = 1$ can contribute.  States with $J>1$ would not
have a nonzero matrix element with the vacuum via the current
${\bar q}\gamma_\mu (1 + \gamma_5) b$.  The possibility of an
intermediate bottom-like meson with $J=0$ is disallowed since it could
only mix with a final state $J=0$ particle and the decay of a spinless
particle to another spinless particle plus a photon is forbidden.

\vspace{0.3in}
The values arrived at in this section should
be considered as upper bounds for the following reason.
We have considered just the lightest possible intermediate states,
because only for these particles is there
sufficient data for making a reasonable phenomenological determination.
However, for the type I amplitude, the kaon intermediate state propagates
far off-shell.  Instead of having a squared momentum near
$q^2 = m_K^2$, the kaon carries $q^2 = m_{B}^2 \gg m_K^2$.
This effect should suppress the transition amplitude by an unknown amount.
In principle, one is to sum over intermediate states.  Contributions
from excited states should be less affected by this
suppression.  Although there is not sufficient data to make a
numerical estimate of their effect, we can anticipate that

\ \ (i) the propagator contribution will indeed be larger, but

\ (ii) the weak-mixing between a ground state $B$ meson and a
radially excited meson $P_n$ will be wave-function suppressed and

(iii) the radiative coupling constant $g_{M \gamma P_n}$ will be
relatively smaller due to phase space competition with other decay
modes of meson $M$.

Qualitatively, the net effect of these considerations would be
expected to decrease the overall radiative amplitude.
\section{\bf Conclusion}

In general, there can be no doubt that any study of long-distance
effects for the heavy-meson transition $B \to K^* + \gamma$ is
a very difficult task.  For example, even a standard technique
such as dispersion theory faces a host of contributing multiparticle
intermediate states, and there exists no rigorous approximation
scheme for dealing with these.  Our feeling is that the most
theoretically and empirically accessible long-distance contribution
is the VMD amplitude, and our study of its quantitative role in
$B \to K^* + \gamma$ is probably the most secure of our results.
Interestingly it turns out to be the largest of the effects that we
considered.  As regards non-VMD contributions, we restricted our attention
to pole diagrams.  It is only for these that we have sufficient
knowledge of the underlying parameters to do the field theory
calculation with any confidence.

The analysis in Sect.~2 of the VMD decay amplitude had two
aspects.  First, there was the demonstration that a gauge
invariant formulation is possible provided that input from the
$B \to V_1 V_2$ transition is restricted to the
transversely polarized part of the $V_1 V_2$ final state.
The phenomenological study which followed indicated a VMD
component in the range
\beq
\Bigg| { {\cal A}^{\rm (VMD)}_{B\to K^* \gamma} \over
{\cal A}_{B\to K^* \gamma}^{\rm (expt)}} \Bigg| \le 0.1 \ \ .
\label{c1}
\eeq
A long-distance contribution to the amplitude of $10\%$ would
imply, via interference, an effect on the decay rate of $20\%$.
The above estimate of the VMD amplitude is smaller than the one
we gave earlier$^{\cite{gp}}$, and it is instructive to see why.
Three different experimental effects each turn out to decrease
the VMD effect:
\begin{enumerate}
\item increase in $B$ lifetime value (from $1.1$~ps to $1.63$~ps),
\item decrease in ${\rm Br}_{B\to K^* \Psi}$ (from $3.6\cdot 10^{-3}$
to $1.73\cdot 10^{-3}$), and
\item newly available polarization data in $B\to K^* \Psi$ which
sharply limits decay into transversely polarized particles.
\end{enumerate}
The first of these decreases the overall $B$ decay rate and affects
all modes equally.  The latter two are specific to the VMD amplitude
and suppress it relative to the experimental signal.  The overall
effect is a reduction of about $15$ in the `transverse'
$B\to K^* \Psi$ decay rate.

Although the VMD contribution obtained here has been inferred from the
polarization data of just one experiment$^{\cite{alam}}$, the errors
are encouragingly small and we expect our phenomenological finding
to be stable.  Indeed, the very recent $CDF$ announcement$^{\cite{cdf}}$
regarding the helicity content in $B \to K^* \Psi$,
\beq
{\Gamma_L \over \Gamma_L + \Gamma_T } = 0.66 \pm 0.10\ ^{+0.08}_{-0.10}
\label{c2}
\eeq
taken with the $CLEO$ value in Table~3 implies the weighted average
\beq
\left\langle {\Gamma_L \over \Gamma_L + \Gamma_T }\right\rangle
= 0.75 \pm 0.08 \ \ .
\label{c2a}
\eeq
This change hardly affects the inequality in Eq.~(\ref{c1}),
raising the right-hand-side to $0.11$.

Suppose we accept at face value our findings regarding the smallness of
long-distance effects in $B \to K^* \gamma$.  What does this imply
for subsequent studies of this decay?  Within the context of the
Standard Model, it emphasizes the dominance of the
electromagnetic-penguin amplitude.  We expect theorists to focus
on reducing remaining uncertainties in the Standard Model prediction
for this process.  Since it will take a while for this
to happen, we recommend prudence in avoiding overly strong claims.
It has been suggested that opportunities exist for detecting the
presence of physics beyond the Standard Model in
$b \to s \gamma$.$^{\cite{new1}}$  We concur, but at the same time
caution that allowance be made for Standard Model uncertainties, {\it e.g.}
like the ones discussed here.  That is, the inclusive branching ratio
sums over exclusive processes, and each of these is (to a greater or
lesser extent) itself subject to the influence of
long-range effects.$^{\cite{long}}$

As for experimental studies, we stress that as valuable as the
$B \to K^* \gamma$ branching ratio determination has become,
polarization studies of the $K^* \gamma$ final state would
yield significant additional information.  The chiral structure of the
EM-penguin operator, up to ${\cal O}(m_{\rm s} /m_{\rm b})$,
predicts in the notation of Eq.~(\ref{a2}) that,
${\overline B} = {\overline C}$.  As a consequence,
the ratio of helicity amplitudes is
\beq
\Bigg| { {\cal M}_{--} \over {\cal M}_{++} } \Bigg| =
{ m_{\rm B}^2 \over m_{K^*}^2 } \ \ .
\label{c3}
\eeq
This is an even firmer prediction of the short-distance
amplitude than is the branching ratio!

We hope that experimental efforts to improve the already impressive
accuracy of the helicity-dependent transition
rates for $B \to K^* \Psi$ will continue.  Of course,
accurate decay-rate and helicity-content determinations
of transitions like $B \to K^* \Psi' , K^* \rho , \dots$~
would likewise be welcome.  Only with such
information could we extend our phenomenological VMD analysis
beyond the one given here.  If the domination of longitudinal
helicities seen in $B \to K^* \Psi$ continues to hold for the
other transitions, we would expect the VMD chain $B \to K^* \Psi
\to K^* \gamma$ studied here to be the largest.  Based on the current
experimental bound for $B \to K^* \Psi'$, the $\Psi'$ VMD amplitude
is estimated to be roughly $0.4$ of the $\Psi$ contribution
and perhaps even smaller.  Since expectations are not bright
for even observing the $B \to K^* \rho$ transition in the immediate
future, neglect of the $\rho$ VMD amplitude appears well justified.

As we pointed out some time ago$^{\cite{gp}}$,
isospin invariance is a consequence (in the spectator model) of
describing the $B\to K^* \gamma$ decay solely in terms of
the short-distance EM penguin amplitude. That is, in this
approximation, the rates for $B^- \to K^{*-} \gamma$ and
${\bar B}^0 \to {\bar K}^{*0} \gamma$ should be equal.  Such
is not the case for all other possible contributions.  For
example, there is a large isospin violation in the system of
$K^* \to K \gamma$ transitions which would be manifested in the
pole contributions of Sect.~3.  Further isospin dependence
might be expected from dynamical interactions between the $b$-quark
and the light antiquark ({\it i.e.} wave function effects).

Two additional radiative transitions of experimental interest
are $B\to \rho \gamma$ and $B\to \omega \gamma$.  From our vantage,
these cannot be analyzed with the phenomenological method described
here because the appropriate data does not yet exist.  Therefore,
purely theoretical models must be employed, and as a result the
predictions for these decays will be rather model dependent.
This work will be described in a separate publication.

Finally, we note that a forthcoming paper will deal with long range effects
in charm meson radiative decays.$^{\cite{bghp}}$  This is especially
interesting because, for charm transitions, the magnitude of the penguin
short-distance contribution is greatly suppressed.  Thus, it should
be possible to experimentally probe the long-distance sector much
more cleanly.  Fortunately, experimental sensitivity is beginning to
reach meaningful levels.$^{\cite{cleo2}}$

The research described in this paper was supported in part by
the National Science Foundation and the Department of Energy.
We wish to acknowledge useful conversations with X. Tata, P. O'Donnell,
G. Burdman, J. Hewett and especially to thank T. Browder for his
continuing interest and encouragement in this project.